\documentclass[conference]{IEEEtran}
\usepackage{amssymb,amsmath,amsthm}
\usepackage{graphicx}
\usepackage{caption}
\usepackage{subcaption}
\usepackage{mdwtab}
\usepackage{soul}
\usepackage{cite}
 \usepackage{multirow}
 \usepackage{color} 
  \usepackage[linesnumbered,ruled]{algorithm2e}
\usepackage{setspace}
\usepackage{esvect}
\usepackage{amsmath}
\usepackage{booktabs}
\usepackage{tabularx}
\usepackage{adjustbox}
\begin{document}
\title{Practical Privacy in WDM Networks with All-Optical Layered Encryption}
\author{\IEEEauthorblockN{Anna Engelmann and Admela Jukan}
\IEEEauthorblockA{Technische Universit\"at Carolo-Wilhelmina zu
Braunschweig, Germany\\
Email: \{a.engelmann, a.jukan\} @tu-bs.de}
}
\maketitle

\begin{abstract}
%\boldmath
Privacy in form of anonymous communication could be comparably both faster and harder to break in optical routers than in today's anonymous IP networks based on The Onion Routing (Tor). Implementing the practical privacy all-optically, however, is not straightforward, as it requires key generation in each anonymization node to avoid distribution of long keys, and layered encryption, both at the optical line rate. Due to the unavailability of cryptographically strong optical key generation and encryption components, not only a layered encryption is a challenge, but an optical encryption in general. In this paper, we address the challenges of optical anonymous networking for the first time from the system's perspective, and discuss options for practical implementation of all-optical layered encryption. To this end, we propose an optical anonymization component realized with the state-of-the-art optical XOR logic and optical Linear Feedback Shift Registers (oLFSRs).  Given that LFSR alone is known for its weak cryptographic security due to its linear properties, we propose an implementation with parallel oLFSRs and analyze the resulting computational security. The results show that proposed optical anonymization component is promising as it can be practically realized to provide a high computational security against deanonymization (privacy) attack.% in optical networks. 
\end{abstract}

\IEEEpeerreviewmaketitle

\section{Introduction}
The onion routing (Tor) is the most popular anonymous routing, implemented as an overlay network among volunteer IP layer systems. Tor deploys the so-called onion routers (ORs) that are communicating using pairwise TCP connections, whereby anonymous communication is possible with layered encryption through traffic tunneling over a chain of ORs. Even though Tor is best known to providing anonymous communication to the end-nodes, also the network providers appreciate the Tor design principles as they can be used to prevent traffic analysis performed by an adversary. In Tor, the original data is encrypted multiple times (layered encryption), and along the path each selected OR decrypts one layer of encryption prior to forwarding, "peeling the onion." While the layered encryption feature makes Tor a de-facto solution for anonymous networking, it creates at the same time the main performance drawback, resulting in an unbalanced distribution of Internet traffic due to tunneling, and a large processing delay in ORs. This motivates our interest in investigating whether anonymous and secure routing can be implemented in the optical layer. For a starter, Internet tunneling can be implemented with wavelength circuits, and the processing performance can be improved with all-optical processing.

\par To support privacy in optical networks, as well as security,  a few challenges need to be addressed from the system's perspective. First, we need to integrate two basic optical components: an optical encryption component and an optical key generator. Second, we need optical layered encryption. Optical encryption device is essential, whereby optical data needs to be logically combined with the secure key. The key generation is also critical, as it needs to implemented in each node to avoid the distribution of a long anonymization keys over the network or large buffers for key storage. The key generation, however,  does not  only need to be realized all-optically, but it must provide a cryptographically secure long key, as well as be performed at line rate. Two state-of-the-art optical components can be used to this end: optical XOR (oXOR) and optical Linear Feedback Shift Registers (oLFSRs). oXOR has been proposed as a practical symmetric key cryptography choice as it can easily concatenate plain text with a key \cite{4, 6, 16, 20}, however has not been used yet for layered encryption. LFSR is simple to implement all-optically based on a shift register and oXORs \cite{6}. Since LFSR is however known for its weak cryptographic security, it cannot be included in the system in a straightforward fashion \cite{9}. 

\par In this paper, we propose a practical solution for all-optical layered encryption and key generation, while addressing the issue of weak cryptographic security of LFSR from the system's perspective. In our system, we utilize an electronic non-linear pseudo random number generator (pRNG) to provide parameters for configuration of an Optical Key Generator (oKG) based on oLFSRs; this is in contrast to the common usage of pRNG which is for the actual key generation in electronics, and not for \emph{configuration of key generation}. To generate a key with a high level of computational security and at line speed, we propose to use multiple parallel oLFSRs, each corresponding to different generator polynomial, and, in addition, to periodically switch between them, \emph{reset} of oKG. We investigate the computational security of proposed system by assuming a deanonymization attack. Based on the security analysis we reverse engineer the  system parameters required for practical realization of oKG such as the oLFSR length, number of parallel LFSRs, maximal key length, number of required resets, line rate adaptation as well as switching time between parallel oLFSRs. The results show that the proposed all-optical layered encryption can be practically implemented and provide high level of computational security. 

The rest of the paper is organized as follows. Section 2 describes the system, along with the anonymization components proposed. System analysis is presented in Section 3. Numerical results are shown in Section 4. Section 5 concludes the paper.

\section{System Design}
%\begin{figure*}[t]
% \centering
%\includegraphics[width= 0.85\textwidth]{Bild2}
%  \caption{All-optical WDM source and destination with encryption capability.}
%\label{net}
%\end{figure*}
\begin{figure*}[!t]
\centering
\includegraphics[width=0.65\textwidth]{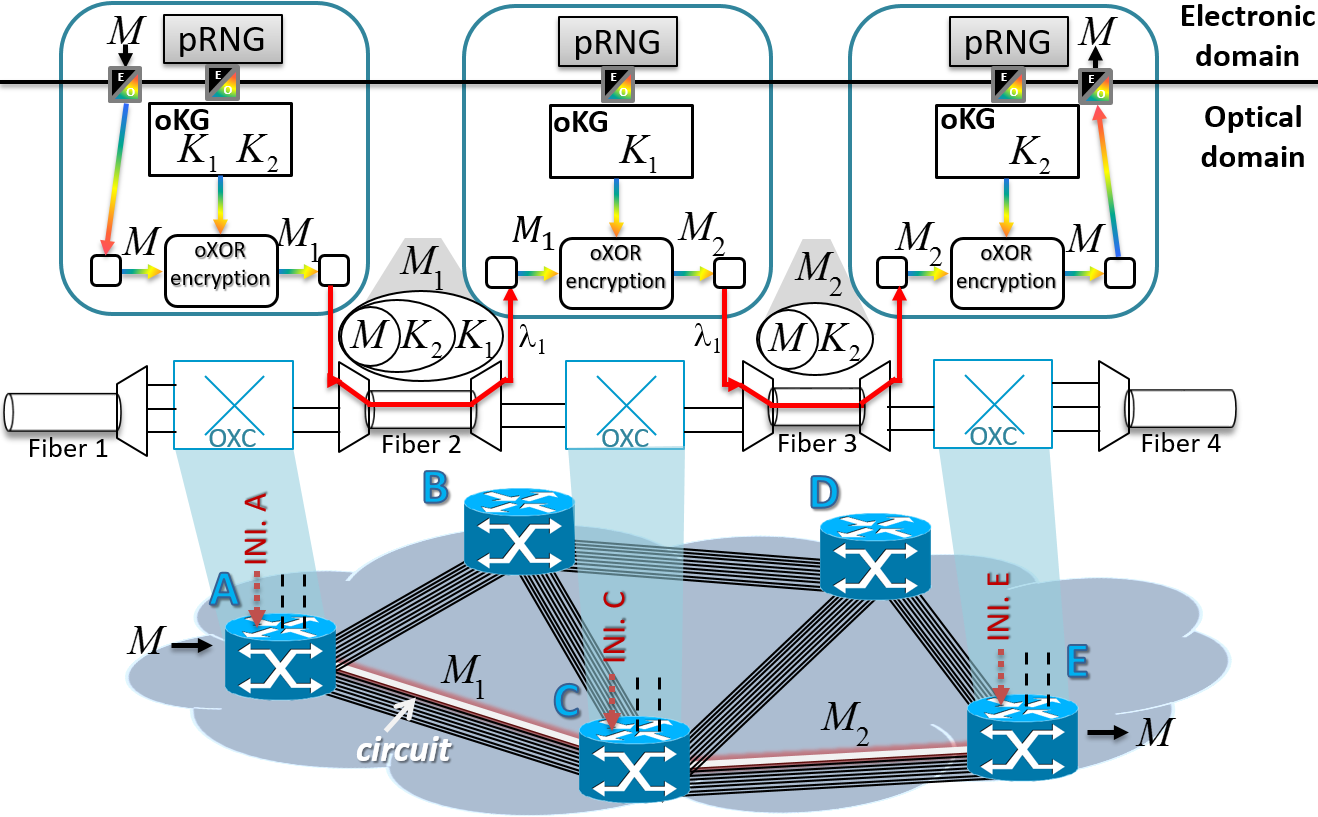}%
  \vspace{-0.1cm}
  \caption{Optical transmission system with layered encryption capability.  $K_i$: key; $M$: original data; $M_i$: data encrypted with $K_i$.}
  \vspace{-0.6cm}
  \label{net}
\end{figure*}

\subsection {Optical "ToR" network}
\par Fig.\ref{net} illustrates an anonymous optical network based on the traditional optical WDM network, with configurable optical cross connects. We assume that the optical control plane is able to assign wavelengths, perform routing and resource allocation, and in addition control the configuration of the anonymization components. The anonymization components are added to the traditional cross connect architecture and include pseudo-random number generator (pRNG), optical key generator (oKG) and optical encryption component based on optical XOR (oXOR) \cite{3,5,8}. The only component implemented in electronics is pRNG, and it requires electro-optical conversion. In our approach, we implement pRNG at the transmission speeds comparably lower than the related optical line rate, and use the output of that pRNG to configure oKG such that a long secure bit sequences at line rate can be produced as a key in each anonymization node on the circuit. Once the pseudo random bit sequence is converted into optical signal, it is interpreted by oKG as \emph{parameters} for generation of anonymization key. This combination of a low rate pRNG and high-bit rate oKG is one of the salient features of our proposed architecture, since cryptographically strong pRNGs at typical optical line rates are not yet available. Once the key is generated, it is combined with the optical data in the optical encryption component. The optical encryption component with oXOR can insert or remove encryption layers by applying anonymization keys generated by oKG. 

%It should be noted that the control plane distributes the start value, i.e., INI. signal, for pRNG utilized in each anonymization node on circuit, whereby pRNG produces random parameters for generation of anonymization key. That key is generated by the oKG, which is start up by initialized pRNG. 

\par Similarly to Tor, we distinguish between source and anonymization nodes. The source node inserts as many encryption layers as there are anonymization nodes along the circuit (in Tor it is 3). The anonymization nodes on the circuit remove one encryption layer each. For illustration, let us consider an example of a circuit setup between nodes A (source) and E (destination), with two hops, and an anonymization node C in between. Once the circuit is established, the control plane distributes the start value, for instance an initialization (INI) signal, for each pRNG utilized on the circuit, whereby pRNG produces random parameters for generation of anonymization key. The original data $M$ in source is layered encrypted with two keys  $K_1$ and $K_2$, i.e., of nodes C and E, by applying oXOR operation. The XOR operation transforms the original data $M$ of length $L_M$ into a new bit sequence $M_i=[M\oplus K_i]$ by applying a key $K_i$ of length $L_K=L_M$ related to anonymization node $i$ and, thus, anonymizes the communication on the circuit. The anonymization in anonymization nodes, i.e., decryption, is implemented by applying the same key $K_i$ with XOR operation as follows $M_i\oplus K_i=[M\oplus K_i\oplus K_i]=M$. In Fig.\ref{net}, the source A encrypts the original data $M$ in layers as $M_1=[M\oplus K_2\oplus K_1]$, whereby nodes C and E remove one encryption layer by decryption. When encrypted optical data $M_1$ reaches the next node C, it is anonymized, i.e., decrypted, with the same key $K_1$ as applied by source, i.e., $[M_1\oplus K_1]=[M\oplus K_2\oplus K_1\oplus K_1]=[M\oplus K_2]=M_2$. In the destination E, encrypted optical data $M_2$ is decrypted with the anonymization key $K_2$ as $[M_2\oplus K_2]=[M\oplus K_2\oplus K_2]=M$, converted to electronic signal and then sent to the higher layers.

%\begin{figure*}[!t]
%\subfigure[All-optical LFSR]{
%\includegraphics[width=0.75\columnwidth]{./fig/AOLFSR.png}
%\label{AOLFSR}
%}\hspace{5mm}
%\subfigure[State machine of T-function]{
%\includegraphics[width=1.2\columnwidth]{./fig/figTF.png}
%\label{figTF}
%}
%\vspace{-0.2cm}
%\caption{Implementation of Key Generator }\label{KG}
%\vspace{-0.3cm}
%\end{figure*}

%\begin{figure}[t]
% \centering
%\includegraphics[width= 0.2\textwidth]{fig/AOLFSR.png}
% \vspace{-0.4cm}
%\caption{All-optical LFSR.}
%  \vspace{-0.4cm}
%\label{AOLFSR}
%\end{figure}

%\IEEEpubidadjcol

\begin{figure}[t]
 \centering
\includegraphics[width= 0.33\textwidth]{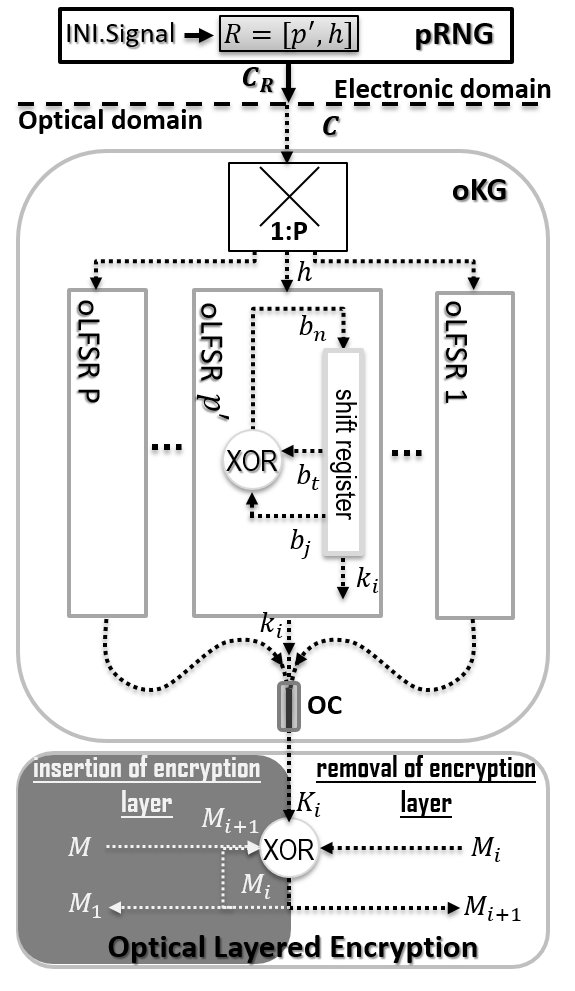}
 \vspace{-0.2cm}
\caption{Anonymization components proposed; $p'$: bit stream to choose oLFSR; $h$: LFSR seed of $n$ bits; $M$: original data; $M_i$: data $M_{i+1}$ encrypted with a key $K_i$ of anonymization node $i$; $C_R$: bit rate of pRNG; $C$: line rate; $P$: number of oLFSRs in oKG; $k_i$: a part of key $K_i$ generated during reset cycle.}
  \vspace{-0.6cm}
\label{ParLFSR}
\end{figure}

\subsection{Anonymization components}

Fig. \ref{ParLFSR} zooms into the main anonymization components proposed, including the oKG based on parallel oLFSRs and the optical layered encryption, based on an oXOR encryption loop.  For these components to work, pRNG from the electronic layer is required to configure oKG in any anonymization node. This is necessary since there is no optical implementation of cryptographically strong pRNGs. Therefore, the bit rate $C_R$ of electronic sequence from pRNG is generally much lower (up to 1.3 Gb/s \cite{18,15}) than the line rate $C$ of optical data. The pRNG generates a bit sequence $R=[p',h]$ of length $L_R$, which we refer to as \emph{true secret key}. The oKG interprets the true secret key $R$ as its parameters for configuration, whereby $p'$ corresponds to the index of an oLFSR to be used, and  $h$ is the seed used by that oLFSR. This simple mechanisms makes it possible to enhance the electronic bit sequence $R$ from pRNG from a short secret key at low bit rates in electronics, into a long anonymization key $K_i$ at optical line rate.

%To implement anonymization components practically, a few challenges need to be addressed. First, an optical encryption component with capability to insert multiple encryption layers needs to be implemented. Second, oKG must be integrated into each optical node to avoid the distribution of a long anonymization keys. However, that oKG must provide a cryptographically secure long key and be realized all-optically to operate at line rate.  Fig. \ref{ParLFSR} shows an architecture of anonymization component, which consists of electronic pRNG, oKG based on oLFSRs for key generation and optical encryption component based on oXOR.

\subsubsection{Optical Key Generator (oKG)}
%To generate a strong anonymization key $K_i$ in each anonymization node utilized, there is a need for two components: a strong secure pRNG at low speed and oKG based on $P$ oLFSRs at line rate. The pRNG generates a bit sequence of length $L_R=[p',h,z]$, which we refer to as \emph{true secret key}. The oKG interprets the true secret key as parameters, i.e., as an index of oLFSR $p'$ and as a seed $h$, while sequence $z$ is optional and can be utilized by oKG to increase randomness of anonymization keys. The seed $h$ is then forwarded by an optical $1:P$ switch to the oLFSR $p'$, whereby optical signal only from that oLFSR $p'$ is forwarded to oXOR gate as an anonymization key. The bit sequence of true secret key $L_R$ from pRNG is, in fact, enhanced by oKG into a long anonymization key $K_i$, whereby key length and bit rate are adapted as required, i.e., to the length of secret data $M$ and to the line speed $C$. 

As previously mentioned, pRNG  generates a \emph{true secret key}, $R$, i.e., index $p'$ to select oLFSR for key generation and the seed $h$ used as an oLFSR start sequence. Since pRNG is initialized with predefined start value (INI.Signal) in all optical nodes along the circuit, pRNGs in the source and any anonymization nodes generate the same pseudo random numbers, i.e., the same parameters ($p'$, $h$) for oKG. Thus, oKG along the routes are synchronized and generate the same anonymization keys $K_i$.

The optical LFSR extends a short input start sequence (seed $h$ from pRNG) into a long output sequence and to operate at optical line rate, e.g., up to 250 Gbit/s \cite{6}, whereby clocking signal implemented by RZ laser at the same speed sends optical signal (a series of logical ones) at line rate to determine the data-rate of the oLFSR \cite{patent:2005}. However, LFSR is known for a weak cryptographic security \cite{17,9}, due to its linear properties. To this end, we propose in this paper two system modifications to increase the computational security, i.e., the time complexity for guessing the anonymization key. The first modification includes an implementation of parallel oLFSRs. The second is the key reset. 
%utilized in source and any anonymization nodes, whereby these keys generated can be \textit{infinitely long}. %Depending on configuration and bit rate $C_R$, pRNG can define new true secret key continuously or periodically and, for instance, can be configured by the control plane.

%The main task of oKG is to extend the short sequence $R$ from pRNG into a long anonymization key $K_i$ as well as to adapt the low bit rate $C_R$ to the line rate $C$. The main element of oKG is oLFSR due to its ability to extend a short input start sequence (seed $h$) into a long output sequence and to operate at optical line rate up to 250 Gbit/s \cite{6}, however, LFSR is known for a weak cryptographic security \cite{17,9}, due to its linear properties. For practical potential of oKG based on oLFSRs (Fig. \ref{ParLFSR}) to be secure, we propose to increase the computational security for guessing the anonymization key through two additional system modifications. 

\par The idea behind parallel oLFSRs related is to have a choice of different generator polynomials, whereby any oLFSR $p'$ for generation of anonymization key is defined randomly by pRNG. oKG can include $P$, $1\leq P \leq P_{max}$, oLFSRs, where $P_{max}=\varphi(2^n-1)/n$ is a maximal number of primitive irreducible polynomials of degree $n$ ($\varphi(\cdot)$ is Euler function). Fig. \ref{ParLFSR} shows an all-optical implementation of oKG based on $P$ oLFSRs, where shift registers have a fixed size of $n$ bits and require seeds $h$ of the same length $n$. From this register, a set of fixed bits denoted as $b_t$ and $b_j$, each corrsonding to the utilized generator polynomial, are XOR concatenated and the resulting bit $b_n$ is fed to the shift register at the last position ($n$), whereby the sequence in the register is 1-bit shifted \cite{16}. %When pRNG generates additional bits ($z$), the bit $b_n$ is XOR concatenated with $z$-bit and the result of this operation, i.e., bit $b'_n$, is put into register. Note, the utilization of sequence $z$ increases the randomness of keys generated by oLFSR and computational complexity of key cracking as we show in Sec. \ref{secure}, however, the exact definition of this sequence and analysis of benefits and disadvantages of its using is out of scope of this paper. 

\par The second proposed modification is reset of oKG, i.e., periodically switching between oLFSRs with its reinitialization by random seed $h$. We refer to the time between two resets as \emph{reset cycle}. During any reset cycle, a randomly selected oLFSR generates only a part $k_i$ of anonymization key $K_i$, whereby $N$ reset cycles, i.e., resets, are required to generate a whole anonymization key $K_i$, i.e., $K_i=[k_i^1, k_i^2, ..., k_i^N]$. Thus, the anonymization key $K_i$ of length $L_K$ is generated by multiple randomly selected oLFSRs one by one during $N$ reset cycles and consists of $N$ key parts $k_i$ of length $L_k$, i.e., $L_K=NL_k$.

\subsubsection{Optical Layered Encryption}
 \par  Typically, layered encryption in Tor is implemented with Advanced Encryption Standard (AES) with a key length $128$ bits \cite{ToR}, which is modified multiple times with Rijndael's key schedule and XOR concatenated with data. To implement layered encryption all-optically, we propose the following method. The original data $M$ in the source is XOR concatenated into $M_{i+1}$ with any anonymization keys of nodes from $r$ to $i+1$, $1\leq i\leq r$, where $r$ is a number of anonymization nodes on a circuit. To insert an additional encryption layer, the encrypted optical data $M_{i+1}$ is fed back to the oXOR gate and encrypted into $M_i$ with a key $K_i$, $M_i=[M_{i+1}\oplus K_i]=[M\oplus K_{r}\oplus ... \oplus K_{i+1}\oplus K_i]$. When all $r$ keys from $K_{r}$ to $K_1$ are applied, the optical data $M_1$ leaves the encryption loop. To remove an encryption layer, optical data $M_i$ arrived at anonymization node $i$, is XOR concatenated with the same key $K_i$ such as in source into optical data $M_{i+1}$ sent to the next anonymization node $i+1$.

% similar to \cite{Guan:2016}

 \subsection{A numerical example}
\begin{figure*}[!t]
\centering
\includegraphics[width=0.85\textwidth]{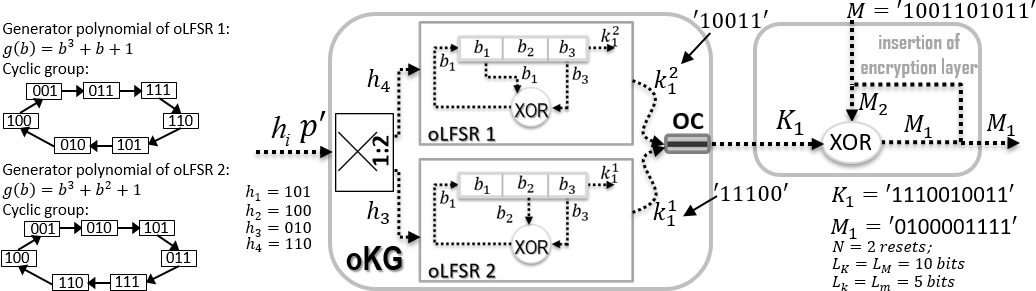}%
  \vspace{-0.2cm}
  \caption{Example of layered optical encryption system with two oLFSRs of length $3$.}
  \vspace{-0.6cm}
  \label{exp}
\end{figure*}
Fig. \ref{exp} shows an example of layered optical encryption, whereby oKG is implemented with two oLFSRs of length $n=3$, i.e., based on two different generator polynomials of degree $3$, i.e., $g(b)=b^3+b+1$ and $g(b)=b^3+b^2+1$. Generator polynomials define bits in shift register for XOR concatenation. Thus, in oLFSR 1, these are bits on the first and last position in register ($b_3$ and $b_1$). Each oLFSR can generate without repeats at most $7$ sequences of maximal length $7$ bits each. Each sequence is defined by a cyclic group, which describes possible seeds and related output sequences. We assume that oKG utilizes the oLFSR1 if it receives bit $p'=$'0' from pRNG, otherwise oLFSR2. The original data $M$ consists of $L_M=10$ bits, while $N=2$ resets are required to generate anonymization key $K_i$ of the same length $L_K=10$ bits, i.e., key consists of $2$ parts $k^1_i$ and $k^2_i$ of $L_k=5$ bits each. Let's assume the circuit of $3$ nodes as presented in Fig. \ref{net}, i.e., $2$ anonymization nodes (C and E) on the circuit. Thus, two encryption layers, i.e., two keys $K_1$ and $K_2$, are required. 

\par When optical data $M$ arrives at the source, pRNG is first initialized by an INI. E signal of destination. oKG receives the first configuration bits $[p',h_1]=$'0101' from pRNG, i.e., $p'=$'0' and seed $h_1=$'101', and writes seed bits in shift register of oLFSR 1. The first generated key part $k^1_2$ is a sequence '00111', while bits related to seeds are always skipped, e.g., bits '101' of seed $h_1$ do not belong to the key sequence $k_2^1$. During the first reset cycle, oKG receives next sequence $[p',h_2]=$'1100' from pRNG and writes seed $h_2=$'100' into shift register of the oLFSR 2, $p'=$'1'. The second part of key is $k_2^2=$'10111' and the resulting anonymization key is $K_2=[k_2^1, k_2^2]=$'0011110111'. The original data $M=$'1001101011' is then XOR concatenated with $K_2$ resulting in encrypted data for the second anonymization node $M_2=$'1010011100', which is fed back to the oXOR gate. At the same time, the oKG generates the second anonymization key $K_1$. The pRNG is now initialized with INI.C value of the first anonymization node C and generates pseudo random bits $[p',h_3,p',h_4]=$'10100110' for both reset cycles.  By utilizing seeds $h_3$='010' and $h_4$='110', the streams generated with oLFSR 2 and then oLFSR 1 are $k_1^1=$'11100' and $k_1^2=$'10011', respectively. The resulting key $K_1=$'1110010011' is XOR concatenated with $M_2$ into data $M_1=$'0100001111' sent to the first anonymization node.

\section{System analysis}

\subsection{Deanonymization attack model}
\par The deanonymization attack is modelled as the attacker's attempt to eavesdrop the optical signal with the goal to both guess the source and destination of the optical circuit (routing), as well as the content of the optical data flow. We assume that the attacker can eavesdrop on any incoming and outgoing optical fibers of an anonymization node $i$. Thus, to correlate input and output port of certain optical flow, and reveal the routing, the attacker needs to guess an anonymization key $K_i$ of node $i$ for certain optical flow, decrypt it and finally compare the deanonymized flow with all outgoing flows of node $i$. To decrypt the data, an attacker needs to guess all $\mathfrak K$ anonymization keys applied, $1\leq\mathfrak K\leq r$, and remove all $\mathfrak K$ encryption layers. We assume that an encryption layer can be only removed, if the optical data of length $L_M$ is decrypted (whole data). We also assume that the attacker knows the structure of the oKG utilized, including the usage of LFSRs (which is a common assumption). Therefore, the attacker can generate anonymization keys by utilizing the correct sequence of LFSRs (set of $p'$), and correct related seeds (set of $h$), i.e., correct bit sequence of true secret key from pRNG. Since we do not consider attacks on control plane and assume control channel as perfectly secure and pRNG as unbreakable, an attacker would not be able to predict input and output sequences of pRNG and must perform a Brute Force Attack (BFA) to guess the true secret key, and finally the original data. 

% study a computational security of proposed key generation with resets in case an attacker already knows the structure of oKG utilized. Thus, the attacker can generate anonymization keys by utilizing the correct sequence of LFSRs (set of $p'$), and correct related seeds (set of $h$), i.e., correct bit sequence of true secret key from pRNG. Since we do not consider attacks on control plane and assume control channel as perfectly secure and nonlinear pRNG as unbreakable, an attacker is not able to predict input and output sequences of pRNG and must perform a Brute Force Attack (BFA) to guess the true secret key. 

\subsection{Security analysis}\label{secure}
Given the attacker model, let us first analyze the time required for removing one encryption layer. To this end, the attacker needs to try all $P$ different LFSRs, or related polynomials for any key part $k_i$ and all $(2^n-1)$ seeds for each tested oLFSR $p'$. To guess the key $K_i$, an attacker must try out all $P^N$ combinations of oLFSRs utilized. Then, the time required to remove one encryption layer is 
\begin{equation}\label{T2}
T^b=P^N\cdot \tau\cdot(2^n-1),
\end{equation}
where $\tau$ is a  time required for decryption of one key, while time for generation of one binary key sequence $K_i$ is set to zero, since an attacker can pre-generate some keys. 

To reveal the path (routing), the incoming optical data need to be encrypted with related anonymization key and compared with all outgoing optical flows. If we neglect the time for comparison, the time required to find out the relation between incoming and outgoing optical data is defined by Eq. \eqref{T2}. Furthermore, the time to find out the whole circuit with $r$ anonymization nodes is defined as $\mathcal T^{L}=r \cdot T^b$.
%\begin{equation}\label{TLink}
%T^{linking}_i=E^i_{in}\cdot c T^b
%\end{equation}
When an attacker reveals the original data $M$, the eavesdropped optical data $M_i$ must be decrypted with all $\mathfrak K$ out of $r$ anonymization keys. Let us consider the worst case scenario, where the attacker knows or is able to define the path or parts of the path during the decryption process, such as when optical data have already traversed $r-\mathfrak K$ out of $r$ anonymization nodes. As a result, the time required to reveal the secret data is 
\begin{equation}\label{Tbreak}
%$\mathcal T^{M}=\mathfrak K \cdot T^b$.
T^{M}=\mathfrak K \cdot T^b
\end{equation}

\subsection{Reverse engineering of the system parameters}\label{Creset}
Given the requirement on high computational security against the deanonymisation attacks, we now derive the required parameters for practical implementation of the proposed system. The parameters we consider include the switching time of oKG, optimal key length $L_k$ and minimal required bit rate of pRNG assuming the minimal number of resets $N$.

%The second challenge is to design oKG in such a way that  We define the ratio between sequences $k_i$ and $R$ as an overhead of pRNG
%\begin{equation}\label{over}
%\Theta = L_{R}/L_k
%\end{equation}

\subsubsection{Switching time in oKG}
The time for switching between parallel oLFSRs must be not larger than the reset cycle $t_{rc}$. Thus, the optical $1:P$ switch in oKG (Fig. \ref{ParLFSR}) must be able to switch to the next oLFSR during the shortest reset cycle, which is defined as a time to generate one part $k_i$ of anonymization key $K_i$ at line rate $C$, i.e., 
 \begin{equation}\label{RC}
t_{rc}=L_k/C
\end{equation}

\subsubsection{Optimal key length}
We define an optimal key length $L_k$ generated by oLFSR during reset cycle $t_{rc}$ as a length that is larger than the true secret key $R$ generated by pRNG during the same time, $L_k>L_{R}$, whereby optimal key $k_i$ is generated with minimal number of resets $N$ and has no repeats of bit sequence. The first condition, $L_k>L_{R}$, ensures that the required bit rate of pRNG is not larger than the line rate (and is always true). The second condition reduces switching speed, while the third condition defines the cryptographic quality of the key sequence. However, since the length $L_M$ of optical data $M$ can be very large (Gbits), we generally allow a short oLFSR length, $n$, which is less than $log_2(L_k+1)$, $L_k<<L_M$. Thus, we allow that oLFSR generates a key part $k_i$ during multiple LFSR cycles with a possible repeat of the same bit sequence. That is because oLFSRs based on primitive irreducible polynomials of degree $n$ can generate at most $2^n-1$ bits without sequence repeat. We now derive the optimal key length to provide high computational security.

The maximal key length $L_k$ generated during a reset cycle is generally a function of the number of implemented oLFSRs $P$ and its length $n$, whereby the time for removing of one encryption layer, $T^b$, defined by Eq.\eqref{T2}, and the time for one decoding try $\tau$ can be set to a very long computational time (years) and defined by the state-of-the-art technology, respectively. Let us modify the equation for $T^b$ to inequation, whereby the left hand side must be larger than or equal to the right hand side of the Eq. \eqref{T2}. Since the key length $L_k$ of a key part $k_i$ generated between two resets is smaller than the length of original data $M$, i.e., $L_k<<L_M$, the number of required resets $N$ corresponds to the ratio between $L_k$ and $L_M $, i.e., 
\begin{equation}\label{resets}
N=L_M/L_k	        
\end{equation}
With Eq. \eqref{resets}, Eq. \eqref{T2} can be modified as $\tfrac{T^b}{\tau(2^n-1)}\geq P^{\tfrac{L_M}{L_k}}$, i.e., $L_M log_2(P)\geq L_k log_2\left(\tfrac{T^b}{\tau(2^n-1)}\right)$, yielding the optimal key length generated during a reset cycle as
\begin{equation}\label{Cond2}
L_k\leq\tfrac{L_Mlog_2(P)}{ log_2\left(\tfrac{T^b}{\tau(2^n-1)}\right)}.
\end{equation}

\subsubsection{pRNG bit rate}
In any reset cycle $t_{rc}$, the switching between oLFSRs can generally lead to key generation interruptions, which happens when the same oLFSR is chosen during two sequent reset cycles. To avoid this, we can initialize one oLFSR at the same time when another oLFSR is generating a key. On the other hand, such system configurations can significantly reduce the space of possible anonymization keys and, thus, negatively impact the computational security defined by Eq. \eqref{T2}, as it follows that $\hat T^b=P\cdot (P-1)^{(N-1)}\cdot \tau\cdot(2^n-1)$. To maintain the high computational security in presence of interruptions, the pRNG bit rate plays a large role. 

\par Generally, the oLFSR can be selected two times successively with probability $\tfrac{1}{P}$ resulting in key generation interruption, while, with probability $\tfrac{P-1}{P}$, the key generation is without interruption. If we have interruptions, an electrical or optical buffer can be beneficial, whereby bit sequence from pRNG is generated and stored during reset cycle $t^1_{rc}=t^1_{R}$, where $t^1_{R}$ is a time for generation of random binary numbers $p'$ (oLFSR) of $log_2(P)$ bits and $h$ (seed) of $n$ bits. During the same reset cycle, $t^1_{rc}=t^1_{L}$, oLFSR should in the worst case write and skip $n$ seed bits into and from shift register, respectively, and, finally, generate a key part $k_i$ of length $L_k$. Thus, pRNG and oLFSR generate $n+log_2(P)$ bits and $2n+L_k$ bits within one reset cycle, respectively. Let us assume a bit rate of pRNG and oLFSR as $C_{R}$ and $C_{L}$, respectively. Thus, the reset cycle is defined as $t^1_{rc}=\tfrac{2n+L_k}{C_{L}}\geq\tfrac{n+log_2(P)}{C_{R}}$ and the minimal required bit rate of pRNG in case of interruption is
\begin{equation}\label{C1}
    C^1_{R}\geq\tfrac{(n+log_2(P))C_{L}}{2n+L_k}
    \end{equation}
\par In the second case, i.e., key generation without interruptions, the period for oLFSR initialization, $t^2_{ini}=t^2_{R}+t^2_{skip}$, i.e, time $t^2_{R}$ for generation of $p'$ and $h$ and time $t^2_{skip}$ for skipping of seed bits by new selected oLFSR $p'$, must be not larger than the period $t^2_{L}$ for key generation by previously selected oLFSR, i.e., $t^2_{ini}\leq t^2_{L}=t^2_{rc}$. Thus, during any reset cycle $t^2_{rc}$, pRNG generates $n+log_2(P)$ bits, while new selected oLFSR skips $n$ bits related to random seed $h$. Simultaneously with initialization of new oLFSR, the previous oLFSR generates a key part of length $L_k$ bits only. Thus, the reset cycle is defined as follows $t^2_{rc}=\tfrac{L_k}{C_{L}}\geq \tfrac{n+log_2(P)}{C_{R}}+\tfrac{n}{C_{L}}$ and the minimal required bit rate of pRNG as
\begin{equation}\label{C2}
    C^2_{R}\geq\tfrac{(n+log_2(P))C_{L}}{L_k-n}
    \end{equation}
Finally, the mean value for required bit rate of pRNG $C_{R}$ can be defined with Eqs. \eqref{C1}, \eqref{C2} and \eqref{Cond2} as 
\begin{equation}\label{pRNGBitrate}
C_R=\tfrac{1}{P}C^1_R+\tfrac{P-1}{P}C^2_R. 
 \end{equation}

\section{Numerical Results}
This section presents a numerical analysis of the proposed system.  We define the optical line rate as $C=C_L=100$ Gb/s, and the length of optical data container (the unit of data to be secured) as $L_M=1.25$ Gbits, akin to the OTN/WDM container standard. We assume, that the attacker has access to an ultra high speed computer such as the supercomputer Aurora at $180$ Petaflops \cite{4}, and set the reference value for one decoding try of $\tau$=$10^{-18}$ sec. According to this assumption an AES key of length $128$ bit and $256$ bit will be cracked after $\approx 10^{13}$ and $\approx 10^{51}$ years, respectively.

\par
\begin{figure}[t]
 \centering
\includegraphics[width= 0.5\textwidth]{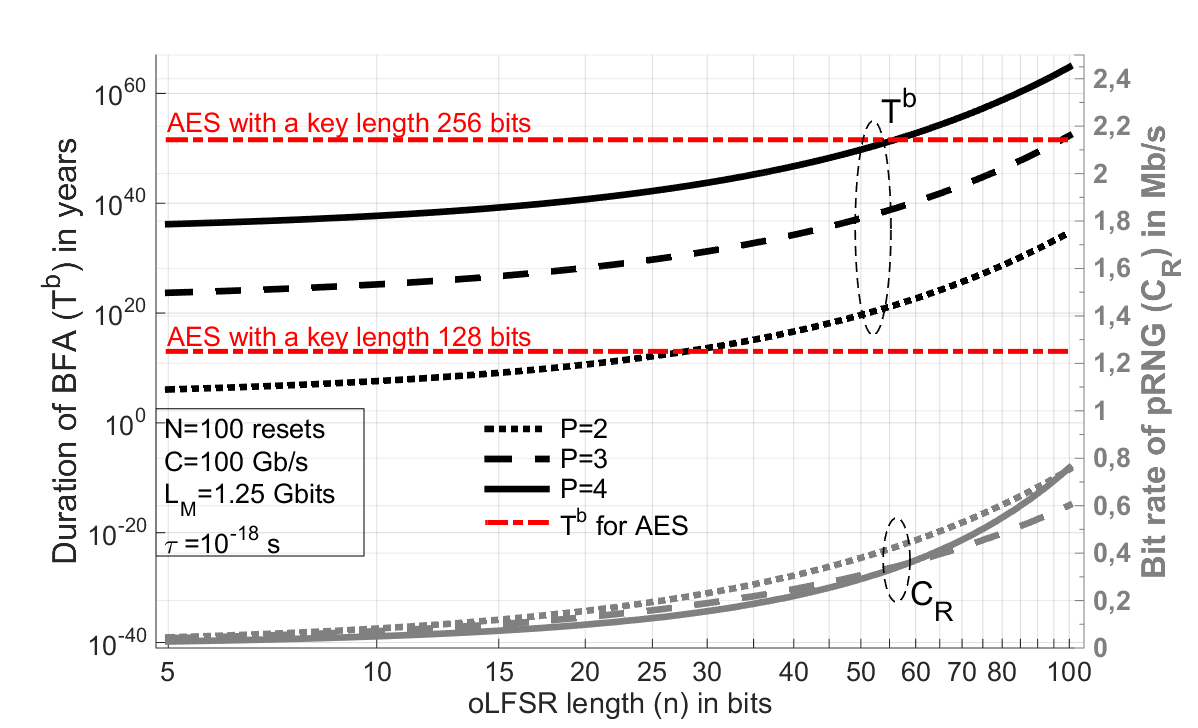}
    \vspace{-0.3cm}
  \caption{Time for BFA and bit rate of pRNG vs. oLFSR length.}
    \vspace{-0.2cm}
\label{fig1}
\end{figure}

 Fig.~\ref{fig1} shows the duration of BFA on encrypted optical flow and required bit rate of pRNG defined by Eqs. \eqref{T2} and \eqref{pRNGBitrate}, respectively, as a function of oLFSR length $n$ and amount of parallel oLFSRs $P$, while the number of resets was set to $N=100$. An increase in length and in amount of oLFSRs increases the security, while the proposed encryption system significantly outperforms the AES 128 in computational security and only oKG based on $P=2$ oLFSRs of length from $5$ to $27$ shows worse performance. The system with $4$ and $3$ oLFSRs of length from $52$ and $98$, respectively, outperforms AES 256 as well. We observe that the proposed oKG can be effectively realized with $3$ oLFSRs of length $n=5$, i.e., $15$ flip-flops, whereby pRNG can operate at much lower bit rate $C_R<0.8$ Mbit/s than optical line rate. Generally, the bit rate of pRNG decreases with increasing number of oLFSRs and decreasing oLFSR length, while the disadvantage is that each oLFSR generates ever longer key part $k_i$, which contains cyclic repeated bit sequences, i.e., $2^n-1<<L_k$. However, that sequence repeat does not reduce the security level. The increasing number of utilized oLFSRs allows to decrease the oLFSR length by keeping the same computational security. For example, oKG based on $3$ OLFSR of length $n=5$ can provide the same time complexity for BFA ($\approx 10^{23}$ years) as oKG based on $2$ oLFSRs of length $n=64$. 

\par We next study the required  switching time of optical $1:P$ switch in oKG. As in case of AES 128, we assume that the duration of BFA on proposed anonymization node must be $\approx 10^{13}$ years. The Table \ref{switch} presents a time required for switching between two oLFSRs defined by Eq. \eqref{RC}. As can be seen, the speed of optical switch can be reduced by increasing the number of oLFSRs ($P$) and its length ($n$).

\par
\begin{figure}[t]
 \centering
\includegraphics[width= 0.5\textwidth]{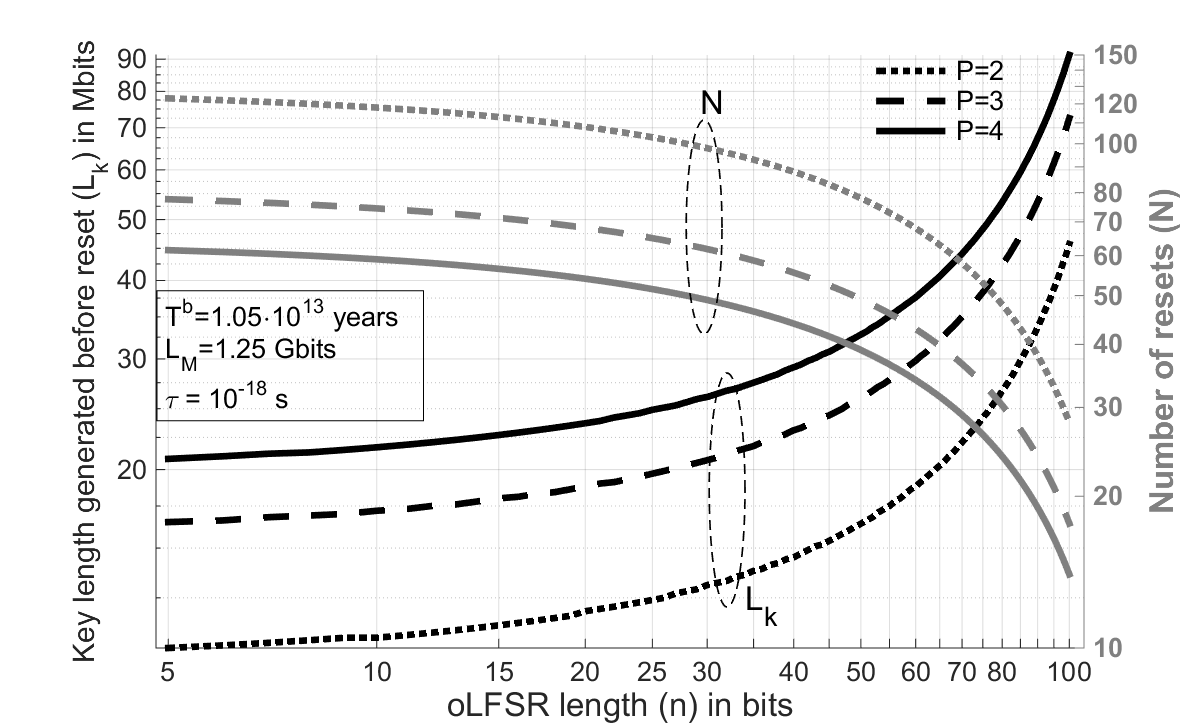}
    \vspace{-0.3cm}
  \caption{Key length and $\#$ of resets vs. oLFSR length.}
  \vspace{-0.4cm}
\label{fig3}
\end{figure}

 Fig. \ref{fig3} shows the maximal key length $L_k$ generated during reset cycle and the required number of resets, per Eqs. \eqref{Cond2} and \eqref{resets}, respectively. With decreasing number of parallel oLFSRs,  $L_k$ also decreases, and increases the number of resets up to $N=123$ for $P=2$ and $n=5$. However, the increasing oLFSR length increases the maximal key length up to $L_K=13.5$ Mbits and decreases resets $N=88$, if $P=2$ and $n=40$. 

\par
\begin{figure}[t]
 \centering
\includegraphics[width= 0.5\textwidth]{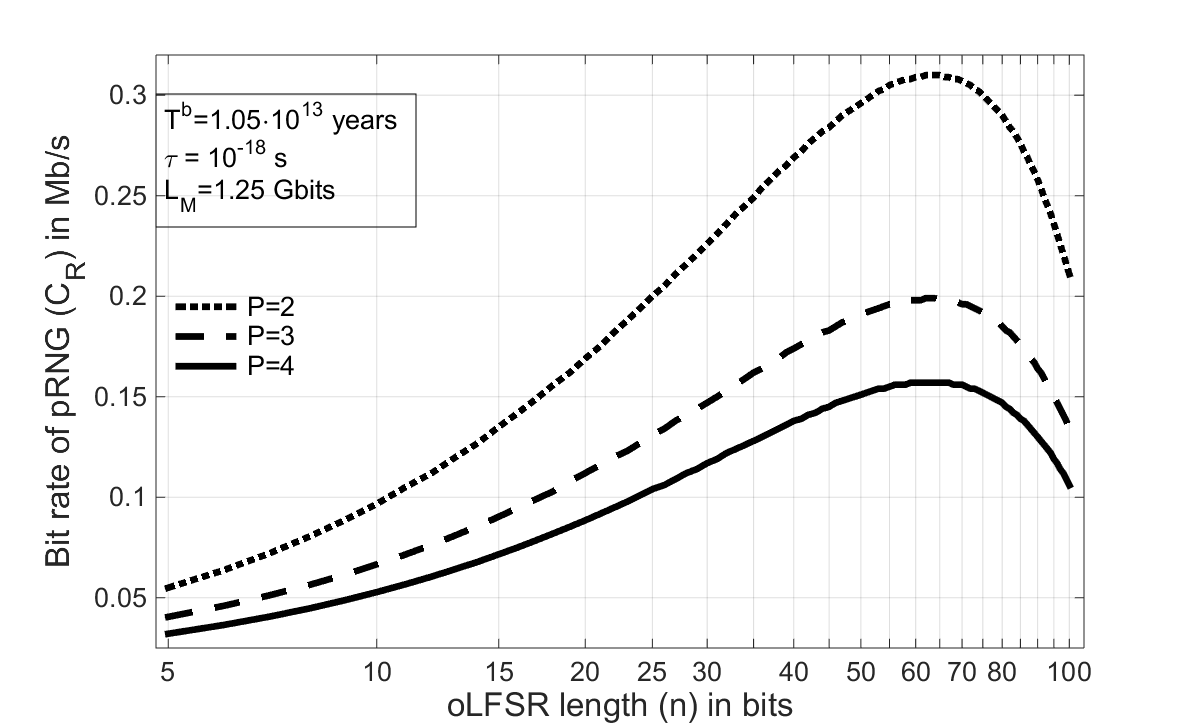}
    \vspace{-0.3cm}
  \caption{Bit rate of pRNG and its overhead vs. oLFSR length.}
  \vspace{-0.1cm}
\label{fig4}
\end{figure}

 We study in Fig. \ref{fig4} the minimal bit rates of pRNG as defined by Eq. \eqref{pRNGBitrate}. The results show that with an increased number of oLFSRs, $P$, and with the decrease in its length $n$, decrease the required bit rate of pRNG $C_R$. For instance, the bit rate of pRNG must be at least $C_{R}\approx0.032$ Mb/s, if the line rate is $C=100$ Gb/s and oKG is implemented with $P=4$ oLFSRs of length $n=5$. We observe that it is practical to utilize pRNG at higher bit rates to generating additional bits, which would in turn further improve the quality of key sequence.

\begin{table} [t]
\begin{center}
\scalebox{0.8} {
\begin{tabular}{cccc}
\hline
\multicolumn{4}{r}{Number of oLFSRs ($P$)} \\
\cline{2-4}
oLFSR length ($n$)    & 2 & 3& 4\\
\hline
5      & 109 $\mu s$  & 173 $\mu s$  &  218 $\mu s$  \\
  10        & 114 $\mu s$        & 180 $\mu s$   &  227  $\mu s$  \\
15       & 119 $\mu s$     & 188 $\mu s$  &   238 $\mu s$ \\
20       & 124 $\mu s$     & 197 $\mu s$    & 249 $\mu s$ \\
\hline
\end{tabular}
}
\caption{Time $t_{rc}$ for switching between oLFSRs.}\label{switch}  
\end{center} 
\vspace{-9mm}
\end{table}


\section{Conclusion}

In this paper, we addressed for the first time the practical challenges of optical anonymous networking from the system's perspective, with new designs of optical key generation and layered encryption at line rate. The results showed that proposed optical anonymization components are promising as it can be practically realized with $3$ parallel oLFSRs of length $5$ operating at $100$ Gb/s to provide a high computational security against deanonymization (privacy) attack, whereby the time for switching between oLFSRs is in the range of microseconds.

%We studied a possible implementation of all-optical layered encryption with the state-of-the-art optical XOR logic and optical LFSRs. We proposed system implementation based on parallel optical LFSRs and analyzed the resulting computational security and required system parameters. The results showed that proposed anonymization system implemented already with $3$ optical LFSRs of length $5$ can provide high computational security against deanonymization (privacy) attack. 

%\appendices

%% use section* for acknowledgment
%\ifCLASSOPTIONcompsoc
%  % The Computer Society usually uses the plural form
%\else
%  % regular IEEE prefers the singular form
%\fi

\ifCLASSOPTIONcaptionsoff
  \newpage
\fi

\bibliographystyle{IEEEtran}
\bibliography{bibL2}
\end{document}